\documentclass[12pt]{llncs}
\usepackage{longtable}
\RequirePackage{graphicx}
\usepackage{epsfig}
\usepackage{multirow}
\usepackage{subfigure}

\newcommand{\ben}{\begin{enumerate}}
\newcommand{\een}{\end{enumerate}}

\usepackage{geometry}
\def\myfigure#1#2{\centerline{\epsfxsize= 0.95 #1\epsffile{#2}}}

\def\myfig2#1#2#3{
\begin{figure}
\centering
\hrule{\hfill}
\myfigure{\columnwidth}{#1}
\caption{#3}
\label{#2}
\hrule{\hfill}
\end{figure}
}

\def\myfig#1#2#3{
\begin{figure}
\centering
\includegraphics[width=2.5in]{#1}
\caption{#3}
\label{#2}
\end{figure}
}

\newcommand{\comment}[1]{\mbox{}}
\newcommand{\meta}[1]{{\bf #1.}}

\title{Dynamic User-Defined Similarity Searching \\ in  Semi-Structured Text Retrieval}

\author{Filippo Geraci\thanks{Affiliated also with the Dipartimento di Ingegneria dell'Informazione, Universit\`a di Siena, Italy. Work
 partially supported  by the Italian Registry of  ccTLD``it"} \and Marco Pellegrini}

\institute{Istituto di Informatica e Telematica del CNR, Via G.
Moruzzi 1, 56100-Pisa (Italy).
\email{filippo.geraci@iit.cnr.it,marco.pellegrini@iit.cnr.it}}

\begin{document}
\maketitle

\begin{abstract}
\noindent
Modern text retrieval systems often  provide a {\em similarity search utility},
that allows the user to find efficiently a fixed number $k$ of
documents in the data set that are {\em most similar} to a given query
(here a query is either a simple sequence of keywords or the identifier of a full document found in previous searches that
is considered of interest).
\noindent
We consider the case of a textual database made of semi-structured documents. For example, in a corpus of bibliographic records
any record may be  structured into three fields: title, authors and abstract, where each field is an unstructured free text.
Each field, in turns, is modelled with a specific vector space. The problem is more complex when we
also allow each such vector space to have an associated {\em user-defined dynamic weight} that influences its
contribution to the overall dynamic aggregated and weighted similarity.
\noindent
This dynamic problem has been tackled in a recent paper by Singitham et al. in \cite{SingithamMR04} in VLDB 2004.
Their proposed solution, which we take as {\em baseline},
is a variant of the cluster-pruning technique that has the potential for scaling
to very large corpora of documents, and is far more efficient than the naive exhaustive search.
\noindent
We  devise an alternative way of embedding weights in the data structure, coupled with a  non-trivial
application of a clustering algorithm based on the {\em furthest point first} heuristic for the metric k-center problem.
The validity of our approach is demonstrated experimentally by showing  significant performance improvements
over the scheme proposed in \cite{SingithamMR04}.
We improve significantly tradeoffs between query time and output quality with respect to the baseline method in
\cite{SingithamMR04}, and also with respect to a novel method by Chierichetti et al. to appear in ACM PODS 2007 \cite{PanconesiRag07}.
We also speed up the pre-processing time by a factor at least thirty.
\end{abstract}

\section{Introduction}

Singitham et al. in \cite{SingithamMR04} consider the following
problem: given set $S$ of $s$ sources of evidence and  a set $E$
of $n$ records, they define for each record $e^j \in E$ and each
source $s_i \in S$ a source score $\sigma_i(e^j)$, moreover for
each source $s_i$ we have a scalar positive weight $w_i$ that is user-defined and
changes dynamically for each query. The dynamic aggregated score of
$e^j$ is $\sum_{i=1}^s w_i\sigma_i(e^j)$. The {\em Dynamic Vector
Score Aggregation} problem is to find the $k$ elements in $E$ with
the highest  dynamic aggregate score. The authors note that in
absence of any further structure the only solution is an
exhaustive computation of  the aggregate score for all the
elements in $E$ and the determination of the $k$ highest elements
in the ranking induced by the aggregation score. Therefore they
consider the special case when each feature of the records $e^j$
is actually a vector, and the source score function
$\sigma_i(e^j)$ is a geometric distance function measuring the
distance of $e^j$ to a query point $q$ (equivalently one can
define a dual similarity function to the same purpose).

\noindent
They also observe that if $s=1$ and the source score is a geometric proximity function 
(e.g. a metric) to a  query point then this problem reduces to the classical $k$-nearest 
neighbor problem. The difficulty in handling the $k$-nearest-neighbor problem in the general 
case of a linear combination of $s \geq 2$ geometric proximity functions stems from the need 
of combining the scores form generally unrelated sources compounded with the presence of 
arbitrary positive weights.
In \cite{SingithamMR04} the Vector Score Aggregation problem is solved by extending the 
cluster pruning technique for the geometric k-nearest-neighbor.

\subsection{Our contribution}

In this paper we use the cluster pruning approach but
we derive  a new and much simpler way of (not) embedding
dynamic weights in vector cluster pruning similarity searches.
Moreover  by using  a different clustering strategies and techniques we obtain further benefits.
In particular we will describe alternatives for the following key
aspects:

\noindent
{\bf (a) How weights are embedded in the scheme}.
In the general Vector Score Aggregation problem the user supplies a query
(this can be either a document in the database or a collection of keywords that capture the concept being searched for) and
a weight-vector that express the user's perception of the relative importance of the document features in capturing the
informal notion of "similarity".
We show in Section \ref{simplif} that, surprisingly, one need not be concerned with dynamic weights at all during
pre-processing, the solution for the unweighted case is good also for the weighted one.

\noindent
{\bf (b) Multiple clusterings.}
In cluster pruning search one decides beforehand to visit a certain number of clusters whose "leaders" are closest to the query point.
However, there is a hidden law of diminishing returns: clusters further away from the query are less likely to contain good k-neighbors.
We use a different strategy: we form not one but several (three in our experiments) different independent clusterings and we search all three of
them but looking into fewer clusters in each clustering.

\noindent
{\bf (c) The ground clustering algorithm}.
When searching for nearest neighbors of a query point $q$ it is natural to consider a cluster good for such a search
when its diameter is small. This leads to considering the optimal K-center problem
(i.e. finding a decomposition minimizing the maximum diameter of any cluster produced)
as a better objective to attain with respect to other conceivable objectives.
Thus we are led to consider the Furthest-Point-First heuristic, that is 2-competitive for this problem \cite{Gonzalez85}.
We attain two benefits: (1) quality of the output is increased, as demonstrated by the experiments in Section \ref{experiment},
(2) preprocessing time is reduced by orders of magnitude since we can use fast variants of this algorithm
(see e.g. \cite{gpsp-sam2-06,gpsm-armil2-06}).


\noindent
By  introducing these three variations we significantly outperform  state of the art algorithms for this problem.

\comment{
In this paper we build on the intuition of \cite{SingithamMR04}
and we bring improvements by proposing  a new way of embedding
weights in vector space based similarity searches. Moreover  by using  a
different clustering techniques we obtain further benefits. By  introducing these two
variations we significantly outperform the original schemes in
\cite{SingithamMR04} for applications in text retrieval. In
particular we will describe alternatives for the following key
aspects:

\noindent
{\bf (a) How weights are embedded in the scheme}.
In the general Vector Score Aggregation problem the user supplies a query document
(this can be either a document in the database or a collection of keywords that capture the concept being searched for) and
a weight-vector that express the user's perception of the relative importance of the document feature in capturing the
informal notion of "similarity". One can easily conceive applications were a fixed weight vector has been found to
be suitable for the application and the user is relieved of the responsability of supplying the weight-vector.
More advanced  applications might benefit from the extra degrees of freedom given to the user in selecting autonomously
the weight-vector.

\noindent
{\bf (b) The ground clustering algorithm}.
The higher the quality of the initial off-line clustering the better the performance of cluster-pruning based searching.
In the context of large text corpora the sheer amount of data to be processed puts severe limits to the class of applicable algorithms.
Those that do not scale to handle efficiently  a number of documents ranging in the hundreds of thousands to millions, and dictionaries of
hundred thousands distinct terms are not suitable to handle large corpora now being collected in Web-based applications.
In practice most papers on this subject use a variant of the k-means algorithm (e.g.\cite{SingithamMR04}).
Here we will use instead a variant of the
furthest-point-first algorithm introduced in \cite{gpsp-sam2-06,gpsm-armil2-06}.
In \cite{gpsp-sam2-06,gpsm-armil2-06} the algorithm was tested
to cluster on-line groups of a few hundred snippets (each web snippets being a mini-document composed of a few dozen terms).
Here we show the effectiveness of this algorithm in handling off-line much larger
corpora where each document contains hundreds of distinct terms.
}

\subsection{Experimental results}

\begin{figure*}[htb!]
\begin{center}
\subfigure[TS1]{\scalebox{.65}{\includegraphics{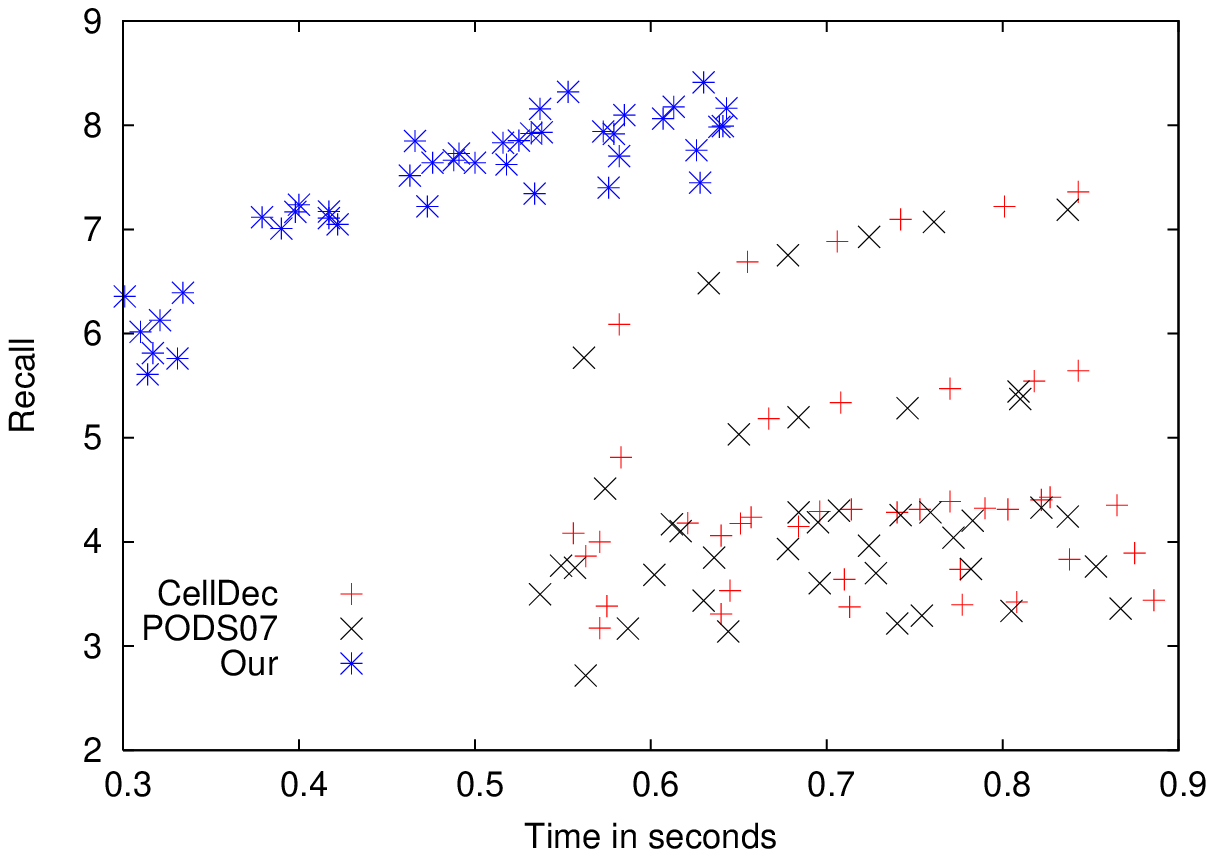}}}
~
\subfigure[TS2]{\scalebox{.65}{\includegraphics{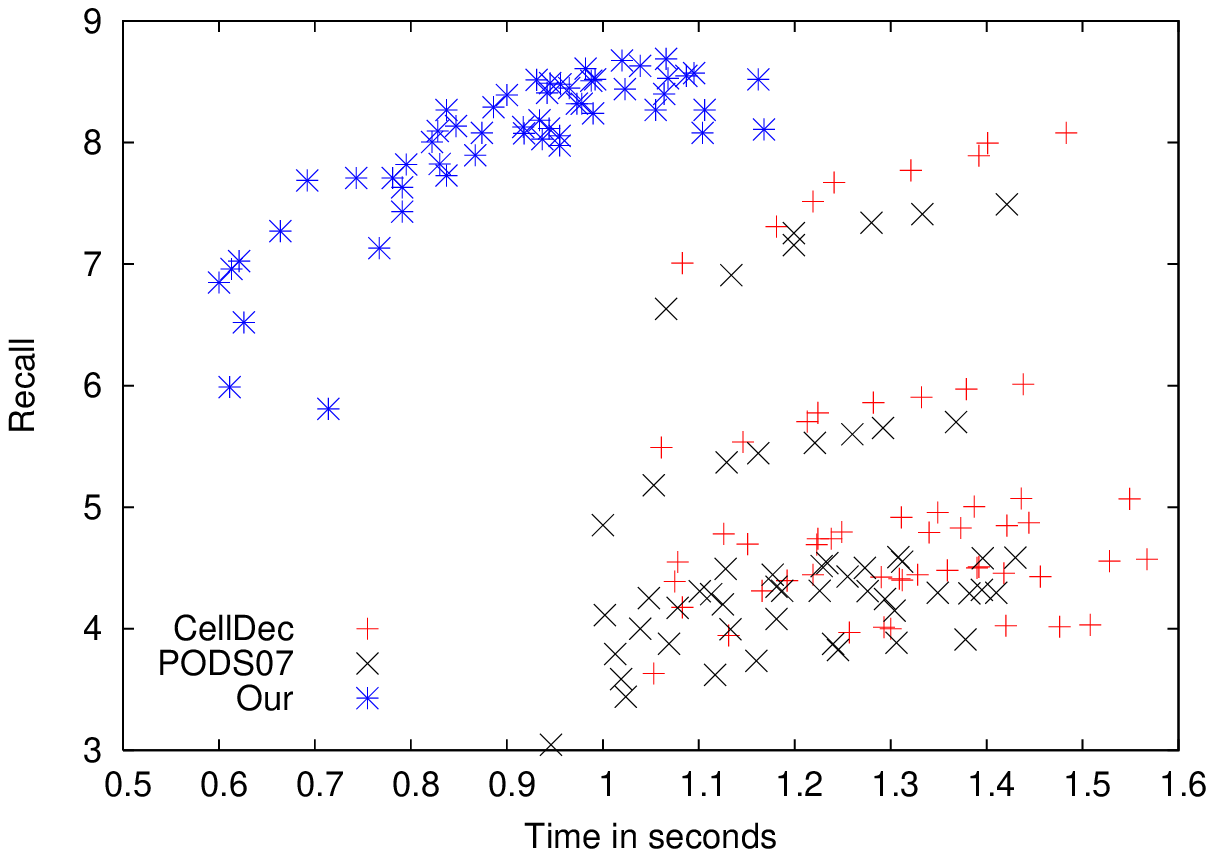}}}
~
\caption{Recall of 10 nearest neighbors as a function of query time.
Each point in the graph is the average of measurements of all queries for a class of weights and a number of visited clusters.
 The points in the upper left corner of the graphs corresponding to our
algorithm show clear dominance.}
\label{fig:tradeoff}
\end{center}
\end{figure*}

We have run our algorithm against two baselines. The first baseline is the algorithm in \cite{SingithamMR04} that uses k-means clustering.
The second base-line is the algorithm in \cite{SingithamMR04} modified so to use a simple cluster pruning randomized strategy proposed in
a forthcoming paper by Chierichetti et al. \cite{PanconesiRag07}. We perform tests on data sets of 50K and 100K documents using a variety
of weights and randomly chosen query documents. Figure \ref{fig:tradeoff} shows the query time/recall tradeoff of the three methods. Our method is
clearly dominant giving consistently better quality results in less time.  Quality data are also given in tabular form in Table \ref{tab:query}.

The top portion of Table \ref{tab:query} corresponds to the case of equal weights, that is equivalent to the unweighted case,  and already
our method shows better time/quality tradeoffs than both the baselines. In the entries of Table \ref{tab:query} for unequal weights our scheme is vastly
superior in recall, even doubling the number of true k-nearest neighbors found using less time over both  baselines.
The overall quality of the retrieved nearest neighbors, as measure via the {\em normalized aggregated goodness},
is also improved: this indicates that our method is robust and stable relative to the baselines.

The simpler clustering strategy in \cite{PanconesiRag07} has preprocessing time close to ours, but quality/cost performance
inferior to our scheme and to that in \cite{SingithamMR04}. Also the improvement in preprocessing time is noteworthy,
we gain a factor 30 against \cite{SingithamMR04} in a test with 100,000 documents.
In practice we could complete the preprocessing in one day compared to one month required by
\cite{SingithamMR04}.

\subsection{Organization of the paper}

This paper is organized as follows.
In Section (\ref{brief}) we give a brief review of the state of the art methods more relevant to our setting,
while a more extended survey is postponed in
the full paper.
In Section (\ref{metrics}) we review known properties of the cosine similarity/distance metric.
In Section (\ref{simplif}) we show the main theoretical analysis underpinning our weight embedding technique.
In Section (\ref{algorithms}) we describe and compare the algorithm that uses our new weight embedding scheme,
and the scheme proposed in \cite{SingithamMR04}.
In Section (\ref{quality}) we describe how the output quality is measured.
In Section (\ref{experiment}) we give the experimental set up and the experimental results.
Conclusions and future work are in Section (\ref{concs}).

\section{Brief state of the Art}
\label{brief}

There is a vast literature  on similarity searching and k-nearest neighbor problems
(see extended surveys in \cite{HjaltasonS03,CNBYMacmcs01}).
However, much less is known for the case when users are allowed to change the underlying metric dynamically
at query time. Besides the  work of \cite{SingithamMR04} we mention work by
P. Ciaccia and M. Patella \cite{CiPa02} discussing which general relations should
hold between two metrics A and B, that allow to build a data structure using the first metric (A),
but perform searches according to the second one (B).

A series of papers by R. Fagin and co-authors \cite{Fag99,FLN03,FaWi00,FaMa00} deal with the problem of rank score aggregation
in a  general setting in which items are ranked independently according to several sources
(not necessarily due to geometric distance functions) and
one seeks to find efficiently  the best combined ranking. The same problem but in a distributed setting is discussed in \cite{MTW05}.
In these papers the rankings are assumed to be statically available and the issue is only how to combine the rankings efficiently. Equivalently,
in their model the cost for producing the independent rankings is not accounted for.  Our setting is different since
the total search cost is considered.

Many other schemes are known for the classical (unweighted, $s=1$) k-nearest neighbor computation
 that, however, are mostly  useful in low dimensional spaces
(or in high dimensional spaces with dense vectors).
Applications in text retrieval are characterized by very high dimensionality
of the corresponding vector space (in the region of tens of thousands) and sparse vector representations.
In general, sophisticated tree-based schemes are ineffective on such data sets.
For example the tree-based algorithm $M^3$-tree proposed by Bustos and Skopal \cite{bustos06}
has been tested on a data set  in dimension 89.
The rank-aggregation method of R. Fagin, R. Kumar and D. Sivakumar \cite{fagin03} has been tested on data sets in dimension  800, while the
hashing based method of  A. Gionis, P. Indyk and R. Motwani \cite{GionisIM99} has been tested data sets in on dimension 64.
The p-sphere method \cite{GoRa00} is used in \cite{PanconesiRag07} as base-line since it has been shown
to be superior to a series of other data structures proposed in literature. Experiments in \cite{PanconesiRag07}
found that their random clustering method performs better than the p-sphere method on textual data.
\section{Metric Spaces and Cosine Similarity}
\label{metrics}

In this paper we will use mostly distance measures, therefore we will convert all results and algorithms
for similarity measures into distances.
As noted in \cite{Cla06} the inner product of two vectors $x$ and $y$ of length 1 (in norm 2) that is the
standard cosine similarity of two normalized vector is turned into a distance $d(x,y) = 1 - x \cdot y$.
This distance function is not a metric in a strict sense since the triangular inequality is not satisfied,
however the following derivation $\|x-y\|_2^2 = x \cdot x + y \cdot y -2 x \cdot y = 2(1- x \cdot y) = 2d(x,y)$
shows that the square root of the distance is indeed a metric. Equivalently one can say that it satisfies the extended
triangular inequality $( d(x,y)^{\alpha} + d(y,z)^{\alpha} \geq d(x,z)^{\alpha})$ with parameter $\alpha = 1/2$.
Moreover a linear combination of distance functions with positive weights defined on the same space is still
a metric space $D(x,y) = \sum_i w_i d_i(x,y)$ for $w_i \geq 0$. Thus the aggregate vector score function used in
\cite{SingithamMR04} although not giving rise to a metric in a strict sense is nonetheless closely related to a metric space.

\section{Simplification of the vector score aggregation problem}
\label{simplif}

In \cite{SingithamMR04} the queries are of the form $q=(q_1,..,q_s)$ where each $q_i$ is a vector of unit length, moreover the user supplies a weight vector
 $w=(w_1,..,w_s)$ where each $w_i$ is a positive scalar weight, and the weights sum to 1.
 The element $e^j$ in the input set $E$ is of the form $(e_1^j,..,e_s^j)$
 where each $e_i^j$ is a  vector of unit length. The aggregate distance function is $d_{AD}(q,e^j) = 1 - \sum_i w_i (q_i \cdot e_i^j)$.
While the aggregate similarity is: $s_{AD}(q,e^j) = 1 - d_{AD}(q,e^j) = \sum_i w_i (q_i \cdot e_i^j)$.

\noindent
\meta{Linearity}
One should notice that because of the linearity of the summation and the inner product operators the weights can be associated to the
vector space: $\sum_i w_i (q_i \cdot e_i^j) = \sum_i  q_i \cdot w_ie_i^j = q \cdot we^j$. This association has been chosen in \cite{SingithamMR04}
thus  the  challenge arises from the  fact that one has to do pre-processing without knowing the
real weights that are supplied on-line at query time.

\noindent
\meta{A different aggregation}
Let $q$ be a query point, $c$ a center of cluster $C(c)$, $p$ a point in cluster $C(c)$, and $D$ a distance function that
satisfies the extended triangular inequality with parameter $\alpha$.
The effectiveness of clustering search stems from the observation that the distance $D(q,p)$ is bounded by an increasing function of $D(q,c)$
and $D(c,p)$. Moreover when $p \in C(c)$, the distance $D(c,p)$ has the smallest value over all  centers in the clustering.
Thus using the center $c$ closest to $q$ gives us the best possible upper estimate of the distance $D(q,p)$.
We have: \[ D(q,p) \leq (D(q,c)^{\alpha} + D(c,p)^{\alpha}))^{1/\alpha} \]

\noindent
Consider now the
{\em  weighted similarity} $WS$:
\[ WS(w,q,p) = \sum_i w_i (p_i \cdot q_i) = \sum_i (w_iq_i) \cdot p_i  = Q_w \cdot p.\]

\noindent
where $Q_w= [w_1q_1,..,w_sq_s]$ is the weighted query vector of vectors.
Since the linear combination of  weights and queries might not result in
a unit length vector we perform a normalization (depending only in the
weights and query point) and obtain a {\em normalized weighted distance} $NWD$:

\[ NWD(w,q,p) = 1- \frac{WS(w,q,p)}{ |Q_w|} =       1-   Q_w/|Q_w| \cdot p     = D(Q'_w,p),        \]

\noindent
where $Q_w/|Q_w| = Q'_w$ is the normalized weighted query vector of vectors.
Now we are in the condition of using the above generalized triangular inequality and establish that:

\[ NWD(w,q,p) = D(Q'_w,p) \leq (D(Q'_w,c)^{\alpha} + D(c,p)^{\alpha}))^{1/\alpha}. \]

\noindent
Since $D(c,p)$ is independent of the pair $q,w$ we can do at preprocessing time
a clustering based on the input set $E$ and the distance $D(.,.)$, regardless of weights and queries.
At query time we can compute $D(Q'_w,c)$ and combine this value with $D(c,p)$ to get the upper estimate of $NWD(w,q,p)$ that guides the searching.
The conclusion of this discussion is that using cosine similarity the multi-dimensional weighted case
can be reduced to a mono-dimensional (i.e. not weighted case) for which we have good data structures.

\section{Algorithms}
\label{algorithms}

\subsection{Basic Cluster Pruning Searching}

For $s=1$ the cluster pruning technique works as follows.
Let $E$ be a set of points in $d$-dimensional space, and $D(.,.)$ the distance function among pairs of points.
The set $E$ is clustered into $K$ groups so to minimize some functional depending on the distance.
Then for each cluster a representative point is elected. When a query point $q$ is given, firs one finds out
a set of $k'$-nearest neighbors among the representative points, for example by exhaustive search. Afterwards only the clusters
whose representative have been selected are searched exhaustively for the $k$-nearest neighbors. All other clusters are not examined,
thus avoiding computing distances from $q$ and the majority of the points in $E$. This procedure is heuristic insofar
as there is no guarantee that all true $k$-nearest neighbors are found, however in practice, by a careful choice of $k$, $k'$ and $K$
one can detect a large fraction of the true $k$-nearest neighbors, while accessing only a small portion of the input data set.

\subsection{Our weight embedding scheme and algorithm}

The discussion in Section (\ref{simplif}) shows that the pre-processing can be done independently
of the user provided weights and that any distance based clustering scheme can be used in principle.
Weights are used to modify directly the input query point and are relevant only for the query procedure.
The basic clustering algorithm we use is described in detail in \cite{gpsm-armil2-06}. It is an algorithm based  on
the further-point-first (FPF) heuristic for the k-center problem that was proposed by \cite{Gonzalez85}.
Summarizing, to produce $K$ clusters we start by taking a sample
of $\sqrt{Kn}$ points out of $n$ points, and we apply the furthest-point-first method on the sample to produce $K$ centers. The
remaining points are associated to the closest center iteratively, while adjusting the representative point (medoid)
of the cluster at each addition of a point to that cluster.

The new twist is that we apply \cite{gpsm-armil2-06} three times on three different random samples
and we collect all the (overlapping) clusters so produced. There is an extra overehead cost in terms of number of distance computations
to be paid at query time when searching multiple clusterings. However, since each our distance
computation  involves only sparse vectors (i.e. we do not use dense centroids), each distance computation is less expensive.
The balance of the two effects is still positive for us as demonstrated by the graphs in figure \ref{fig:time}.

\subsection{The weight embedding scheme of \cite{SingithamMR04}}

In \cite{SingithamMR04} several schemes and variants are compared but experiments show that
the best performance is consistently attained by  Query Algorithm 3 ({\em CellDec})  described in  \cite[Section 5.4]{SingithamMR04}.
The preprocessing is as follows.
For simplicity we consider the 3-dimensional case, that is a data set where each record has 3 distinct sources of evidence
(e.g. in out tests, title, author and abstract of a paper).
We consider the set $T$ of positive weight-vectors summing to one
(this is the intersection of the hyperplane $w_1+w_2+w_3 =1$ with the positive coordinate octant). We split $T$ into
4 regular triangles $T_1$, $T_2$ and $T_3$ each incident to a  vertex of $T$ and the central region $T_4$.
Let $V_i^j$ be the vector corresponding to record $j$ and source $i$.
Region $T_4$ is the central one and weights in $T_4$ are not too different form each other, therefore we form a composite vector
as follows $V(T_4)^j = V_1^j + V_2^j + V_3^j$. For the other four regions we apply a squeeze factor $\theta$ for the vector spaces
corresponding to the lower weights. $V(T_1)^j = V_1^j + \theta V_2^j + \theta V_3^j$. We build similarly $V(T_2)^j$ and $V(T_3)^j$.
Experiments in \cite{SingithamMR04} show that a value of $\theta = 0.5$ attains the best results.
At query time, given the query $Q=(q,w)$ one fist detects the region of $T$ containing $w$, then uses $q$ in the associated indexing
data structure for cluster-pruning.

\comment{
\subsection{Basic clustering: k-center vs k-means}

Apparently it seems that it would suffice to run a k-center algorithm
on the input data (instead of k-means) and the trick is done.
The situation is slightly more complex.
Generally speaking k-means is much slower than k-centers, but produces higher quality clusters (the iterative process has a smoothing effect).

In a different setting, when strictly on-the-fly computations are required, such as those for interactive web snippets clustering
(see \cite{gpsm-armil2-06}\cite{gpsp-sam2-06})
one would set a time limit (e.g. a few seconds) and choose the algorithm giving better quality within this time limit.
In our setting, however, clustering is a pre-computation  done off-line and, while one would appreciate a result in hours rather than days,
clearly, quality of the outcome has to be considered in absolute terms.

We have found that, while one application of k-center is not able to beat
k-means in quality, a few independent applications of randomized k-center produce a set of overlapping clusters that yield better output quality.
There is an extra cost in terms of number of distance computations to be paid at query time when searching multiple clusterings.
However, since each distance
computation in k-center involves only sparse vectors (i.e. there are no dense centroids),
the two effects balance out in query time.
In our experiments three applications of k-center to different random samples of the input suffice.
Note that while we need in our scheme three applications
of k-center, in \cite{SingithamMR04} one needs 4 applications of k-means to build up the structure,
so the gap in preprocessing time is even increased,
while quality improves.
}

\subsection{The clustering scheme of \cite{PanconesiRag07}}

In  a forthcoming paper \cite{PanconesiRag07} Chierichetti et al. propose a very simple but effective scheme
for doing approximate k-nearest neighbor search for documents. In a nutshell, after mapping $n$ documents
in a vector space  they choose randomly $K=\sqrt{n}$ such documents as representatives, and associated each other document to its closest
representative. Afterwards, for each group the {\em centroid} is computed as "leader" of the group to be used during the search.
In \cite{PanconesiRag07} the authors are able to prove probabilistic bounds on the size of each group
which is an important parameter that directly influences the time complexity of the cluster prune search.
Dynamically weighted queries are not treated in \cite{PanconesiRag07},
therefore we choose as a second base-line  to employ \cite{PanconesiRag07},
in place of K-means, within the weighting framework of \cite{SingithamMR04}.
\section{Measuring Output Quality}
\label{quality}

We compare the results provided by the three algorithms using
two quality indexes  (employed also in \cite{SingithamMR04} and
\cite{PanconesiRag07}): the {\em mean competitive recall}  and the {\em mean normalized
aggregate goodness}.

\noindent
{\bf Mean Competitive Recall.}
Let $k$ be the number of similar documents we want to find (in our experiments k=10)
and $A(k, q, E)$ the
set of the k retrieved document by algorithm $A$ on data set $E$, and $GT(k, q, E)$
the "Ground truth", the set of the $k$ closest points in $E$ to the query $q$
which is found through an exhaustive search; the competitive recall is $CR(A,q,k) = |A(k, q, E) \cap GT(k, q, E)|$.
Note that the Competitive recall is a real number in the range $[0,..,k]$ and a higher value indicated higher quality.
The Mean Competitive Recall $\bar{CR}$ is the average of the competitive recall over a set of queries $Q$:
\[ \overline{CR}(A,Q,E) = \frac{1}{|Q|}\sum_{q \in Q} CR(A,q,k) \]

\noindent
This measure tell us how many of the true k nearest neighbors are our algorithm is able to find.

\comment{
\subsection{Mean Aggregate Goodness}
Let $\mu$ be a distance measure, for a query $q$, algorithm $A$ and data-set $E$, the Aggregate Goodness $AG(k,q,A)$ is

\[AG(k,q,A) = \frac{\sum_{p \in GT(k, q, E)} \mu(q,p)}{ \sum_{p \in A(k, q, E)} \mu(q,p)}. \]

\noindent
Note that the Aggregate goodness is a real number in the range $[0,..,1]$ and a higher value indicated higher quality.
The Mean Aggregate Goodness $\bar{AG}$ is the average of the aggregate goodness over a set of queries $Q$:
\[ \overline{AG}(A,Q,E) = \frac{1}{|Q|}\sum_{q \in Q} AG(A,q,k). \]

\noindent
This measure tells us how much of the sum of distances  realized by true k nearest neighbors is close to the sum of distances
of the points found by the algorithm. In other words we are interested in estimating how many $(1+\epsilon)$-approximate l-nearest
neighbors our algorithm is able to find.
}

\noindent
{\bf Mean Normalized Aggregate Goodness.}
We define as the Farthest Set $FT(k,q,E)$ the set of k points in $E$ farthest from $q$. Let the sum of distances of the $k$ furthest points from $q$
be $W(k,q,E)= \sum_{p \in FS(k, q, E)} \mu(q,p)$.
The normalized aggregate goodness:
\[NAG(k,q,A) = \frac{ W(k,q,E) - \sum_{p \in A(k, q, E)} \mu(q,p)}{W(k,q,E) -  \sum_{p \in GT(k, q, E)} \mu(q,p) }. \]

\noindent
Note that the Normalized Aggregate goodness is a real number in the range $[0,..,1]$ and a higher value indicated higher quality.
The Mean Normalized Aggregate Goodness $\overline{NAG}$ is the average of the normalized aggregate goodness over a set of queries $Q$:
\[ \overline{NAG}(A,Q,E) = \frac{1}{|Q|}\sum_{q \in Q} NAG(A,q,k). \]

\noindent
Among the possible distance functions there is a large variability in behavior (for example some distance functions are bounded, some are not).
Moreover for a given $E$ and $q$  there could be  very different range of possible distance values. To filter out all these distortion effects
we normalize the outcome of the algorithm against the ground truth by considering the shift against the $k$ worst possible results.
This normalization allows us a finer appreciation of the different algorithms by factoring out distance idiosyncratic or border effects.

\section{Experiment}
\label{experiment}

In our experiment we compared the following algorithms:

\ben \item[A)] The {\em CellDec} algorithm described in
\cite{SingithamMR04} with k-means
  clustering and weighted cosine distance.
\item[B)] The algorithm proposed in \cite{PanconesiRag07} based on random
  cluster algorithm and weighted cosine distance,
  christened {\em PODS07} for lack of a better name.
\item[C)] The algorithm proposed here based on the
furthest point first algorithm and weighted cosine distance (referred to as {\em Our}).
\een

\noindent
We implemented all the algorithms in Python. Data were
stored in textual \emph{bsd} databases. Tests have been run on a
Intel(R) Pentium(R) D CPU 3.20GHz with 4GB of RAM and with
operating System Linux.

\noindent
Following \cite{SingithamMR04} we have downloaded the first one hundred
thousands Citeseer bibliographic records\footnote{{\tt http://citeseer.ist.psu.edu/}}. Each record contains three
fields: paper title, authors and abstract.
We built two data sets: TS1 with the first 53722 documents and TS2 with all
100000 downloaded documents. After applying standard stemming and stop words
removal, three vector spaces were created: one for each field of the
documents. Terms in the vector are weighted according to the standard {\em tf-idf}
schema. Details are in Table \ref{tab:proprocess}.

\begin{figure*}[htb!]
\begin{center}
\subfigure[Data Set TS1]{\scalebox{.65}{\includegraphics{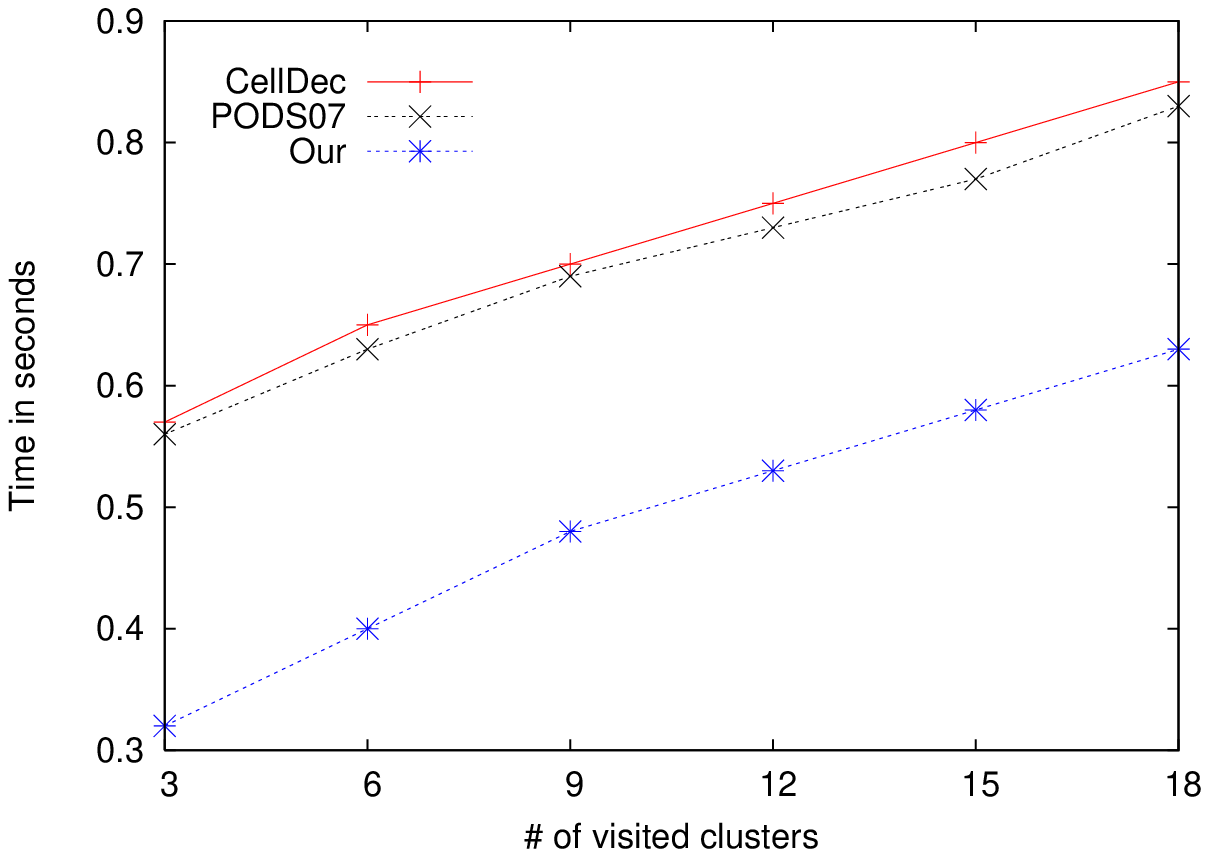}}}
~
\subfigure[Data Set TS2]{\scalebox{.65}{\includegraphics{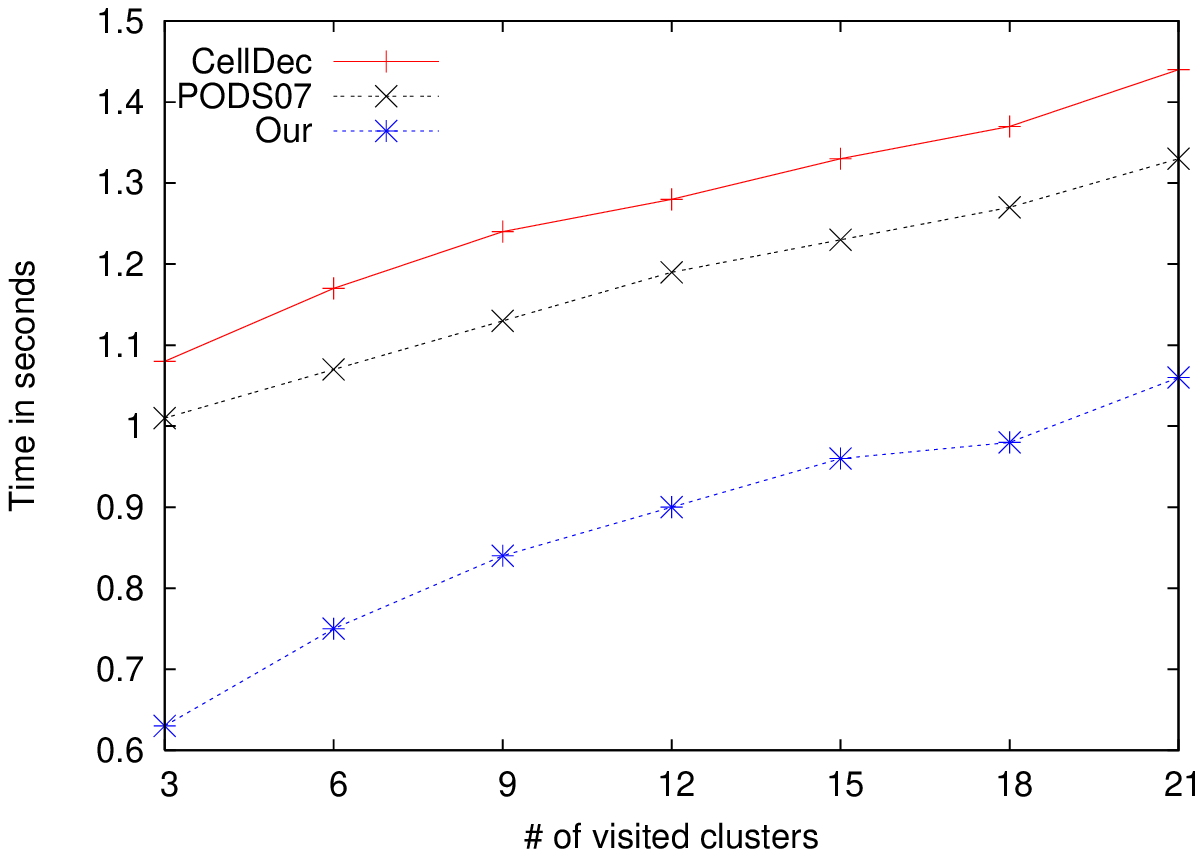}}}
~
\caption{Average query time (in seconds) over all queries in function of the number of visited clusters.}
\label{fig:time}
\end{center}
\end{figure*}

\noindent
Without loss of generality, as queries we used documents extracted from the
data set. Test queries have been selected by picking a random set of 250
documents. During searches  the exact match of the query document is not counted. In our experiments we
adopted the 7 sets of weights used in \cite{SingithamMR04}. For each set of
weights, we always used the same query set. This gave us the opportunity of
comparing results for different choices of the weights vector.
The query time as a function of the number of clusters visited is in Figure \ref{fig:time} and shows clearly the
speed up factor of two.

\begin{table}[h]
  \centering
  \begin{tabular}{|l|c|c|c||c|c|c|}
    \hline
    Dataset & \multicolumn{3}{|c||}{TS1} & \multicolumn{3}{|c|}{TS2} \\
    \hline
    Input size (MB) & \multicolumn{3}{|l||}{41.80} & \multicolumn{3}{|l|}{76.13} \\
    \# Records & \multicolumn{3}{|l||}{53722} & \multicolumn{3}{|l|}{100000}\\
    \# Clusters & \multicolumn{3}{|l||}{500} & \multicolumn{3}{|l|}{1000}\\
    \hline
    Algorithm & Our & CellDec & PODS07 & Our & CellDec & PODS07\\
    \hline
    Preprocessing time & 5:28 & 215:48 & 7:18 & 20:13 & 636.80 & 22:56 \\
    Space (MB) & 332.078 & 1407.656 & 1402.140 & 645.765 & 2738.671 & 2725.078 \\
    \hline
  \end{tabular}
  \caption{Measure of input complexity. Preprocessing time (in hours and minutes) and storage (in Megabytes)
   of the two data structures generated by  {\em CellDec}, PODS07 and our algorithm.}
  \label{tab:proprocess}
\end{table}

\begin{table}[ht]
  \centering
  \begin{tabular}{|l|l||l|l|l|l|l|l||l|l|l|l|l|l|l|}
    \hline
    \multicolumn{2}{|c||}{} & \multicolumn{6}{|c|}{Data Set TS1 = 50K docs.} & \multicolumn{7}{|c|}{Data Set TS2 = 100K docs.} \\
    \hline
    \multicolumn{2}{|c||}{Visited clusters} & 3 & 6 & 9 & 12 & 15 & 18 & 3 & 6 & 9 & 12 & 15 & 18 & 21 \\
    \hline
    \multicolumn{2}{|c||}{} & \multicolumn{13}{|c|}{Query weights 0.33-0.33-0.34 - CellDec weights 1-1-1} \\
    \hline
    \hline
    \multirow{3}{*}{Recall} & CellDec & 6.088 & 6.688 & 6.884 & 7.096 & 7.22  & 7.36  &
                                       7.008 & 7.308 & 7.516 & 7.672 & 7.772 & 7.892 & 7.996 \\
                           & PODS07  & 5.768 & 6.484 & 6.752 & 6.928 & 7.072 & 7.188 &
                                       6.044 & 6.632 & 6.908 & 7.156 & 7.256 & 7.34  & 7.412 \\
                           & Our     & 6.016 & 7.172 & 7.64  & 7.852 & 7.94  & 7.992 &
                                       6.884 & 7.688 & 8.096 & 8.292 & 8.408 & 8.508 & 8.528 \\
    \hline
    \multirow{3}{*}{NAG}   & CellDec & 0.779 & 0.822 & 0.841 & 0.854 & 0.865 & 0.876 &
                                       0.858 & 0.876 & 0.891 & 0.905 & 0.914 & 0.919 & 0.924  \\
                           & PODS07  & 0.753 & 0.816 & 0.831 & 0.842 & 0.852 & 0.863 &
                                       0.779 & 0.827 & 0.851 & 0.867 & 0.874 & 0.881 & 0.887  \\
                           & Our     & 0.776 & 0.838 & 0.863 & 0.876 & 0.879 & 0.882 &
                                       0.842 & 0.887 & 0.907 & 0.915 & 0.921 & 0.925 & 0.927  \\
    \hline
    \hline
    \multicolumn{2}{|c||}{} & \multicolumn{13}{|c|}{Query weights 0.4-0.4-0.2 - CellDec weights 1-1-1} \\
    \hline
    \hline
   \multirow{3}{*}{Recall} & CellDec & 4.812 & 5.184 & 5.336 & 5.472 & 5.544 & 5.644 &
                                       5.492 & 5.536 & 5.704 & 5.776 & 5.86  & 5.904 & 5.972  \\
                           & PODS07  & 4.512 & 5.032 & 5.196 & 5.284 & 5.372 & 5.444 &
                                       4.852 & 5.18  & 5.368 & 5.444 & 5.528 & 5.6   & 5.652  \\
                           & Our     & 6.128 & 7.168 & 7.64  & 7.832 & 7.916 & 7.984 &
                                       6.848 & 7.708 & 8.08  & 8.268 & 8.392 & 8.448 & 8.48   \\
    \hline
    \multirow{3}{*}{NAG}   & CellDec & 0.769 & 0.811 & 0.830 & 0.844 & 0.855 & 0.866 &
                                       0.852 & 0.869 & 0.884 & 0.899 & 0.908 & 0.914 & 0.918  \\
                           & PODS07  & 0.743 & 0.807 & 0.821 & 0.832 & 0.842 & 0.853 &
                                       0.771 & 0.819 & 0.843 & 0.860 & 0.867 & 0.875 & 0.881  \\
                           & Our     & 0.778 & 0.833 & 0.856 & 0.869 & 0.872 & 0.875 &
                                       0.836 & 0.883 & 0.903 & 0.909 & 0.916 & 0.919 & 0.921  \\
    \hline
    \hline
    \multicolumn{2}{|c||}{} & \multicolumn{13}{|c|}{Query weights 0.2-0.4-0.4 - CellDec weights 1-1-1} \\
    \hline
    \hline
    \multirow{3}{*}{Recall} & CellDec & 3.864 & 4.06  & 4.148 & 4.284 & 4.312 & 4.404  &
                                       4.78  & 4.692 & 4.796 & 4.916 & 4.956 & 5.004 & 5.072  \\
                           & PODS07  & 3.772 & 4.168 & 4.284 & 4.3   & 4.284 & 4.328  &
                                       4.0   & 4.2   & 4.344 & 4.428 & 4.5   & 4.552 & 4.58   \\
                           & Our     & 6.356 & 7.116 & 7.516 & 7.624 & 7.704 & 7.76   &
                                       6.96  & 7.708 & 8.004 & 8.076 & 8.184 & 8.24  & 8.268  \\
    \hline
    \multirow{3}{*}{NAG}   & CellDec & 0.698 & 0.737 & 0.756 & 0.774 & 0.786 & 0.797  &
                                       0.798 & 0.811 & 0.828 & 0.847 & 0.857 & 0.863 & 0.868  \\
                           & PODS07  & 0.679 & 0.738 & 0.753 & 0.763 & 0.772 & 0.783  &
                                       0.725 & 0.763 & 0.785 & 0.800 & 0.808 & 0.817 & 0.825  \\
                           & Our     & 0.762 & 0.807 & 0.827 & 0.836 & 0.840 & 0.842  &
                                       0.819 & 0.870 & 0.883 & 0.887 & 0.896 & 0.898 & 0.900  \\
    \hline
    \hline
    \multicolumn{2}{|c||}{} & \multicolumn{13}{|c|}{Query weights 0.4-0.2-0.4 -CellDec weights 1-1-1} \\
    \hline
    \hline
    \multirow{3}{*}{Recall} & CellDec & 4.0   & 4.176 & 4.292 & 4.312 & 4.324 & 4.352 &
                                       4.388 & 4.396 & 4.444 & 4.4   & 4.412 & 4.444 & 4.456  \\
                           & PODS07  & 3.752 & 4.104 & 4.188 & 4.256 & 4.204 & 4.244 &
                                       3.792 & 4.172 & 4.284 & 4.312 & 4.312 & 4.308 & 4.296  \\
                           & Our     & 5.608 & 7.048 & 7.664 & 7.932 & 8.096 & 8.176 &
                                       5.988 & 7.272 & 7.82  & 8.136 & 8.44  & 8.516 & 8.608  \\
    \hline
    \multirow{3}{*}{NAG}   & CellDec & 0.791 & 0.830 & 0.845 & 0.851 & 0.858 & 0.865 &
                                       0.834 & 0.855 & 0.863 & 0.868 & 0.874 & 0.877 & 0.880  \\
                           & PODS07  & 0.757 & 0.815 & 0.828 & 0.839 & 0.849 & 0.856 &
                                       0.762 & 0.813 & 0.834 & 0.848 & 0.852 & 0.855 & 0.858  \\
                           & Our     & 0.786 & 0.869 & 0.901 & 0.916 & 0.922 & 0.926 &
                                       0.817 & 0.895 & 0.924 & 0.934 & 0.943 & 0.946 & 0.949  \\
    \hline
    \hline
    \multicolumn{2}{|c||}{} & \multicolumn{13}{|c|}{Query weights 0.2-0.6-0.2 - CellDec weights 0.5-1-0.5} \\
    \hline
    \hline
    \multirow{3}{*}{Recall} & CellDec & 4.084 & 4.18  & 4.236 & 4.312 & 4.388 & 4.428 &
                                       4.548 & 4.696 & 4.74  & 4.74  & 4.792 & 4.828 & 4.848 \\
                           & PODS07  & 3.496 & 3.684 & 3.848 & 3.932 & 3.964 & 4.04  &
                                       4.112 & 4.252 & 4.308 & 4.444 & 4.492 & 4.516 & 4.54  \\
                           & Our     & 6.392 & 7.008 & 7.22  & 7.344 & 7.4   & 7.448 &
                                       7.024 & 7.632 & 7.824 & 7.976 & 8.028 & 8.056 & 8.08  \\
    \hline
    \multirow{3}{*}{NAG}   & CellDec & 0.770 & 0.802 & 0.818 & 0.828 & 0.842 & 0.848 &
                                       0.870 & 0.900 & 0.914 & 0.923 & 0.927 & 0.928 & 0.930 \\
                           & PODS07  & 0.668 & 0.702 & 0.734 & 0.751 & 0.762 & 0.769 &
                                       0.770 & 0.795 & 0.816 & 0.835 & 0.844 & 0.852 & 0.859 \\
                           & Our     & 0.740 & 0.775 & 0.788 & 0.799 & 0.801 & 0.805 &
                                       0.814 & 0.849 & 0.861 & 0.867 & 0.873 & 0.876 & 0.878 \\
    \hline
    \hline
    \multicolumn{2}{|c||}{} & \multicolumn{13}{|c|}{Query weights 0.6-0.2-0.2 - CellDec weights 1-0.5-0.5} \\
    \hline
    \hline
    \multirow{3}{*}{Recall} & CellDec & 3.172 & 3.308 & 3.376 & 3.396 & 3.424 & 3.44  &
                                       3.632 & 3.944 & 3.968 & 4.012 & 4.0   & 4.024 & 4.016 \\
                           & PODS07  & 2.716 & 3.14  & 3.216 & 3.292 & 3.336 & 3.36  &
                                       3.044 & 3.44  & 3.62  & 3.736 & 3.824 & 3.876 & 3.884 \\
                           & Our     & 5.76  & 7.236 & 7.848 & 8.156 & 8.32  & 8.412 &
                                       5.808 & 7.132 & 7.728 & 8.128 & 8.32  & 8.488 & 8.632 \\
    \hline
    \multirow{3}{*}{NAG}   & CellDec & 0.809 & 0.845 & 0.861 & 0.867 & 0.870 & 0.874 &
                                       0.803 & 0.852 & 0.861 & 0.865 & 0.869 & 0.874 & 0.874 \\
                           & PODS07  & 0.725 & 0.793 & 0.823 & 0.839 & 0.849 & 0.856 &
                                       0.702 & 0.784 & 0.823 & 0.836 & 0.849 & 0.860 & 0.862 \\
                           & Our     & 0.795 & 0.883 & 0.913 & 0.930 & 0.936 & 0.939 &
                                       0.812 & 0.891 & 0.921 & 0.936 & 0.945 & 0.953 & 0.957 \\
    \hline
    \hline
    \multicolumn{2}{|c||}{} & \multicolumn{13}{|c|}{Query weights 0.2-0.2-0.6 - CellDec weights 0.5-0.5-1} \\
    \hline
    \hline
    \multirow{3}{*}{Recall} & CellDec & 3.384 & 3.532 & 3.64  & 3.736 & 3.832 & 3.892 &
                                       4.176 & 4.312 & 4.424 & 4.48  & 4.5   & 4.508 & 4.556 \\
                           & PODS07  & 3.168 & 3.436 & 3.604 & 3.7   & 3.74  & 3.764 &
                                       3.584 & 3.876 & 3.996 & 4.08  & 4.148 & 4.244 & 4.292 \\
                           & Our     & 5.812 & 7.108 & 7.728 & 7.92  & 8.064 & 8.164 &
                                       6.52  & 7.432 & 7.896 & 8.116 & 8.32  & 8.4   & 8.52  \\
    \hline
    \multirow{3}{*}{NAG}   & CellDec & 0.773 & 0.806 & 0.828 & 0.840 & 0.856 & 0.866 &
                                       0.853 & 0.869 & 0.884 & 0.889 & 0.894 & 0.896 & 0.903 \\
                           & PODS07  & 0.737 & 0.785 & 0.812 & 0.825 & 0.835 & 0.840 &
                                       0.755 & 0.807 & 0.834 & 0.845 & 0.852 & 0.862 & 0.866 \\
                           & Our     & 0.773 & 0.859 & 0.887 & 0.896 & 0.902 & 0.908 &
                                       0.837 & 0.889 & 0.914 & 0.923 & 0.933 & 0.936 & 0.939 \\
    \hline
  \end{tabular}
  \caption{Quality results of the compared algorithms. Recall is a number in [0,10], Normalized Aggregated Goodness is a number in [0,1].
  Data as a function of the number of visited clusters.}
  \label{tab:query}
\end{table}


\section{Conclusions and future work}
\label{concs}

We have shown that a difficult searching problem with dynamically chosen weights  can be reduced,
thanks to the linearity properties of the cosine similarity metric, to a simpler static search
problem. For this problem  we provide efficient and effective method that are competitive with state of the art techniques for
large semi-structured textual databases. We plan in future work to extend and test our techniques to handling other types of data
(e.g. images, and sound). We wish to thank P. Raghavan for introducing us to cluster pruning techniques and A. Panconesi
for many useful discussions and for providing a preprint of \cite{PanconesiRag07}.



\end{document}